# Evolution of the Online Rating Platform Data Structures and its Implications for Recommender Systems

Hao Wang
Ratidar Technologies LLC
Beijing, China
haow85@live.com

*Abstract*— Online rating platform represents the new trend of online cultural and commercial goods consumption. The user rating data on such platforms are foods for recommender system algorithms. Understanding the evolution pattern and its underlying mechanism is the key to understand the structures of input data for recommender systems. Prior research on input data analysis for recommender systems is quite limited, with a notable exception in 2018 [1]. In this paper, we take advantage of Poisson Process to analyze the evolution mechanism of the input data structures. We discover that homogeneous Poisson Process could not capture the mechanism of user rating behavior on online rating platforms, and inhomogeneous Poisson Process is compatible with the formation process.

*Keywords—recommender system, homogeneous Poisson Process, inhomogeneous Poisson Process*

## I. Introduction

Online rating platform was born together with the internet. Amazon.com is one of the earliest online product review platforms that not only allows users to purchase goods online but also gives consumers opportunities to rate their purchased products. With the emergence of Web 2.0, social network sites such as Douban.com, Goodreads, and new versions of e-commerce websites such as Temu have been rapidly developing into new battlefields whose data are coveted by scientists and engineers. The availability of data led to the invention of big data technologies. Hadoop and Spark raised a tsunami in the internet industry and machine learning algorithms have become so widely acceptable that every big internet company has at least one team specialized in machine learning related technologies.

Most scientists focus on improving the accuracy of recommender systems. Greatest inventions in the field are mostly accuracy boosters such as Factorization Machines and DeepFM. Although the major driving force behind the technology is the commercial desire to increase traffic volume and decrease the marketing cost. Other concerns have raised awareness among industrial workers in the last ten years. Fairness is one such topic that leads to global government policy remediation. Fairness issues include demographic bias, popularity bias, selection bias, position bias, etc. We are still at a very early stage of fairness understanding, and in the future years, as machines become more intelligent, the issue would raise more awareness among the public.

Another issue related to recommender system is the cold-start problem. Many zeroshot learning approaches have been invented for the problem. The de facto standard paradigm for the problem is transfer learning / meta learning with pretrained model. However, in recent years, a new school of zeroshot learning algorithms that require no input data sources emerged as a better alternative. The representative algorithms include ZeroMat [1], DotMat [2], PowerMat [3] and PoissonMat [4]. The cultural and social implications of these approaches are also explained in [5].

Although recommender system have enticed tens of thousands of scientists into the field. Only a small handful of researchers have been focusing on data analysis. In 2018, Wang [6] published a paper quantifying the impact of the popularity bias in the input data structures on the output data structures as well as the intermediate computational steps. This paper was the first of its kind in the field that provides precise analytical formulas that capture the combinatorial mechanism underlying the collaborative filtering approaches.

In this paper, we investigate into the evolution mechanism underlying the formation of the online rating platform dataset. We discover that although Poisson Process could capture the user rating behavior properly, homogeneous Poisson Process is not compatible with the formation process, and inhomogeneous Poisson Process not only could capture the formation process, but also have interesting properties that we'd like to discuss in details in the following sections.

II. Related Work

Recommendation is an important task for internet product workflows. The simplest and earliest recommender system is collaborative filtering [7][8] and its variants [9][10]. Collaborative filtering assume that people with similar historic preferences have similar product choices in the future. Unlike collaborative filtering, matrix factorization is a dimensionality reduction and data completion algorithm that has dominated the field for a decade. Important inventions of this paradigm include Probabilistic Matrix Factorization [11], SVDFeature [12], SVD++ [13], timeSVD [14] and ALS [15]. In recent years, scientists have used matrix factorization to solve fairness problem [16][17][18], cold start problem [1][2][3][4], and context-aware recommendation problem [19][20].

As deep neural networks started to boom in 2010's, complex machine learning models such as Wide & Deep [21], DeepFM [22], DLRM [23], AutoRec [24] an AutoInt [25] have emerged as important techniques in the field. The main advantage of the apporaches is the accuracy improvement, and the major drawback is the slow speed that usually requires GPU machines.

Although scientists and engineers have published extensively on the algorithmic design aspect of the field, it is still very difficult to find research publications on data analysis and algorithm verification topics. One notable work in the field is Wang [6], which quantifies the popularity bias effect of the collaborative filtering approaches.

In this paper, we investigate into the formation mechanism of the online rating datasets, and provide insights for recommender system designers. To the best of our knowledge, we are one of the very few data analysis and algorithm verification publications in the field of recommender systems.

III. Formation Process as Homogeneous Poisson Process

As illustrated in PoissonMat [4], the user rating behavior on the online rating platform could be properly modeled using Poisson Process and Zipf Distribution. We model the user rating behavior in the following way :

$$P(R_{i,j} = k) = \frac{\lambda^k}{k!} e^{-\lambda}$$

By Zipf Law, we have :

$$\frac{P(R_{i,j} = k)}{P(R_{i,j} = j)} = \frac{k}{j}$$

In other words, we have :

$$\frac{\lambda^k}{\lambda^j} \times \frac{j!}{k!} = \frac{k}{j}$$

Assume k > j , and solve for $\lambda$ , we have :

$$\lambda = \exp\left(\frac{\ln(k!) - \ln(j!) + \ln(k) - \ln(j)}{k - j}\right)$$

This one equation is not compatible with multiple equations that form an equation system for different (k,j) pairs, and therefore is invalid. Therefore, we use a different formuation of Poisson Process :

$$P(N(t_k + t_0) - N(t_0) = k) = \frac{(\lambda t_k)^k}{k!} e^{-\lambda t_k}$$

By Zipf Law, we have :

$$\frac{P(N(t_k + t_0) - N(t_0) = k)}{P(N(t_j + t_0) - N(t_0) = j)} = \frac{k}{j}$$

Assume k > j, the initial time $t_0 = 0$, and solve for $\lambda$, we have :

$$\lambda t_k = k \ln(\lambda t_k) - \ln(k! \, k)$$

Solving the equation systems with different values of k leads to complex variable solutions, so we draw the conclusion that homogeneous Poisson Process is not compatible with the formation process of online user rating behaviors.

To solve the incompatibility problem with homogeneous Poisson Process, we resort to inhomogeneous Poisson Process.

IV. Formation Process as Inhomogeneous Poisson Process

We use inhomogeneous Poisson Process to build the following equation system with constraints to tackle the problem :

$$P(N(t_k + t_0) - N(t_0) = k) = \frac{(\lambda_k t_k)^k}{k!} e^{-\lambda_k t_k}$$

By Zipf Law, we have the following equation :

$$\frac{P(N(t_k + t_0) - N(t_0) = k)}{P(N(t_j + t_0) - N(t_0) = j)} = \frac{k}{j}$$

Expanding the equation system with constraints, we have :

$$(\lambda_j t_j - \lambda_k t_k)(k \log \lambda_k - j \log \lambda_j + k \log t_k - j \log t_j + \log j! - \log k!) = \log k - \log j$$
$$t_j < t_k, \; if \; j < k$$

We solve the equation system with linear constraints using Sequential Least Square Programming [26], and obtain the following plots :

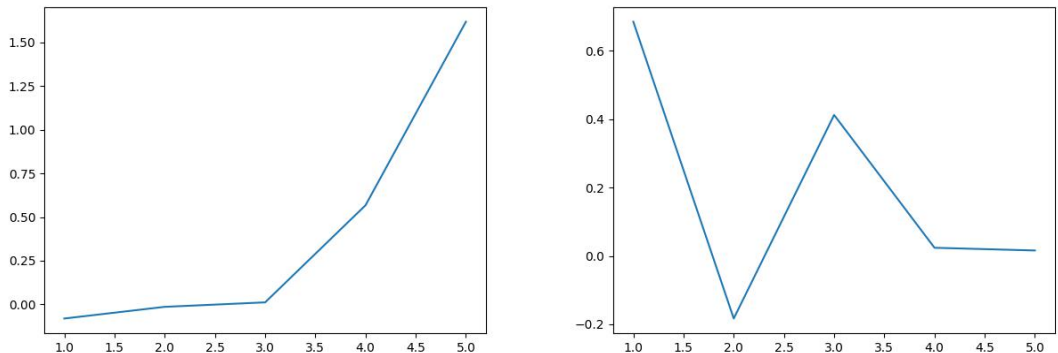

Fig. 1 A solution to the inhomogeneous Poisson Process equation systems : Different time values for rating data (left) ; different $\lambda$ values for rating data (right)

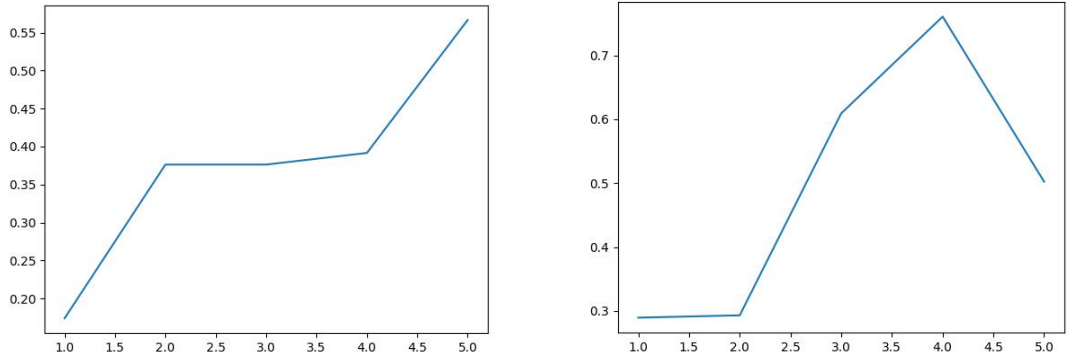

Fig. 2 A solution to the inhomogeneous Poisson Process equation systems : Differnt time values for rating data (left) ; different λ values for rating data (right)

Fig. 1 and Fig. 2 results are obtained from 2 different runs of Sequential Least Squares Programming. Due to the instability of the numerical methods, we could not draw deterministic conclusions on the properties of the time variables and λ values. However, one certain thing is, we could generate experimental results that are compatible with the assumptions. In this case, inhomogeneous Poisson Process is a compatible model for the formation process of user item rating behavior dataset.

Please notice the PoissonMat model [4] is an inhomogeneous Poisson Process model, and it is capable of producing competitive results with other modern day techniques. The model is therefore justified by our statistical modeling results.

## V. Discussion

By our experiments, we notice that the Zipf Law properties of the user item rating behavior for recommender systems could be explained by Inhomogeneous Poisson Process. Therefore, we have built the connection between Pareto Distribution and the Poisson Process. We would like to point out the connection is not only limited to the field of recommender system, but can also be generalized to other areas of machine learning and social sciences.

It would be interesting to point out that by verifying that inhomogeneous Poisson Process is compatible with Zipf Law properties of the online rating platform data formation process, we are able to justify the validity of algorithms such as PoissonMat.

## VI. Conclusion

In this paper, we quantify the formation proces of the online rating platform rating dataset as an inhomogeneous Poisson Process and verifies the theory using experiments. In the process, we not only create a theoretical model for user behavior for recommender systems, but also justifies the theory of PoissonMat, which is a zeroshot nonparametric machine learning model for recommender systems.

In future work, we would like to find a better and more precise optimization method other than Sequential Least Squares Programming, so we could pin down the exact and precise parameters in our experiments so we would be able to explore the properties of our statistical models accurately.